# The Nondipole Part of the Main Geomagnetic Field and the Large Scale Topographical Heterogeneities of the Core-Mantle Boundary

Demina I., Ivanov S., Merkuryev S.

St-Petersburg Filial of Pushkov Institute of Terrestrial Magnetism, Ionosphere and Radio wave Propagation (SPbF IZMIRAN)
`dim@izmiran.spb.ru`

**Abstract**
Based on satellite data, both large and small-scale flows were identified on the core surface .At the same time, the observed rapid decrease in the dipole component of the main geomagnetic field does not find a satisfactory explanation.
In the previous papers we carried out the comparative studies of the structure of the lowest mantle with the motion path of the small size current loops approximating the MGF's small-scale anomalies. The hypothesis was stated that heterogeneities in the lowest mantle structure and the topographic irregularities of the core-mantle boundary associated with ancient subduction zones is one of the primarily responsible for the small-scale vortices formation and the local MGF variation.
In this paper, we carry out the analogous comparison for the large scale sources approximating the nondipole part of the geomagnetic field.

## Introduction

By now, great advances have been achieved modeling processes occurring in the thickness of the Earth's liquid core [1]. Based on satellite data, both large and small-scale flows were identifiedon its surface [2, 3].At the same time, the observed rapid decrease in the dipole component of the main geomagnetic field (MGF) that is attended with the development of its nondipole part does not find a satisfactory explanation.
In [4,5] we carried out the comparative studies of the structure of the lowest mantle with the motion path of the small size current loops approximating the MGF's small-scale anomalies. The hypothesis was posited in [4,5] that heterogeneities in the lowest mantle structure and the topographic irregularities of the core-mantle boundary associated with ancient subduction zones is one of the primarily responsible for the small-scale vortices formation and the local MGF variation.
In this paper, we carry out the analogous comparison for the large scale sources approximating the nondipole part of the geomagnetic field.

## Data

The most powerful (world) anomalies of the MGF's nondipole part were approximated by the field of current systems, which geometry is a hollow thin-walled truncated cone. To calculate the MGF components, we used the spherical harmonic coefficients of the COV-OBS model [6] for the year 2010. To reduce the contribution of small-scale anomalies, the MGF components were calculated at a distance of 10000 km from the center of the Earth. At the greater distance, some world anomalies start to decrease rapidly, what complicates the analysis. The spatial distribution of the nondipole radial Z component used in this work is shown in Fig.1

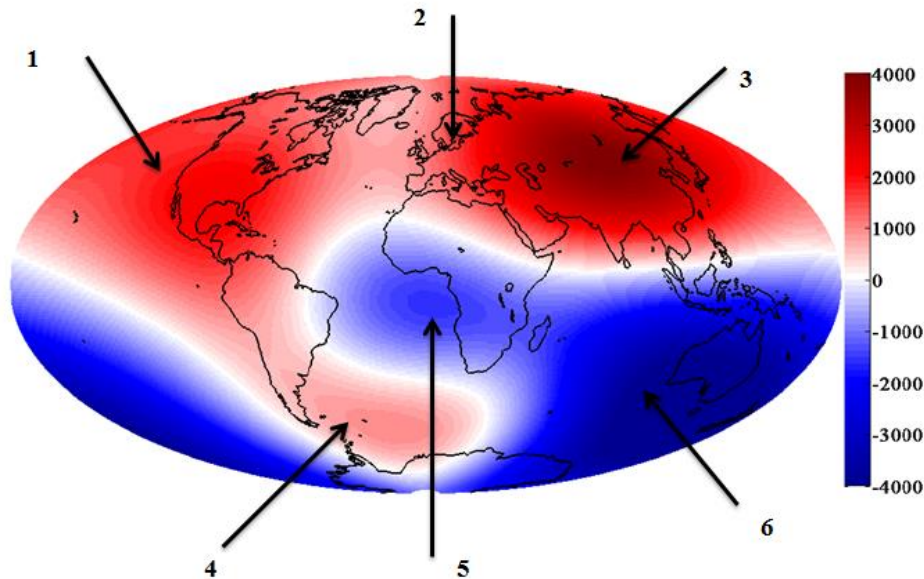

**Fig.1** Spatial distribution of the nondipole radial *Z* component at a distance of 10000 km from the center of the Earth. World anomalies are marked with arrows and numbers.

As well as in previous works, model SAW642AN [7] was used as a model of the structure of the lowest mantle. This model allows us to construct the distribution of the enhanced and reduced propagation velocities of seismic waves relative to the average values for a given depth over the entire mantle, including the core-mantle boundary (Fig. 2). We used this mantle slice for comparison the seismic wave non-uniformities with the position and orientation of the cones, obtained by us from magnetic data. The interpretation of our results is based on the idea, that is not the subject of discussion among experts in seismic tomography, namely, that the regions of higher and lower seismic waves correspond to higher and lower densities of the mantle matter.

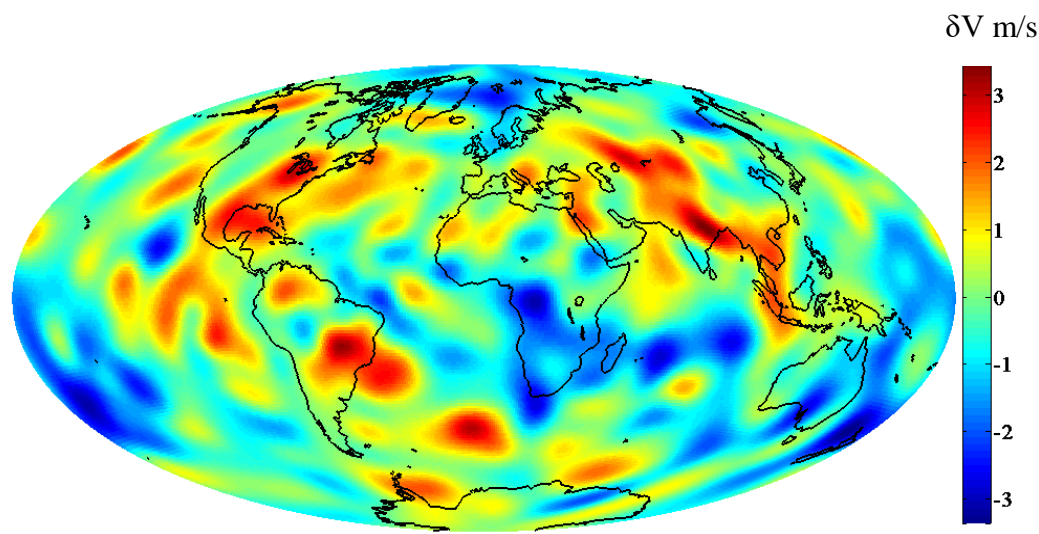

**Fig. 2** Model SAW642 at the core-mantle boundary.

**Method.**
The most powerful anomalies of the MGF's nondipole part marked in Fig.1 were approximated by fields of the volume current systems (VCS), the geometry of which is a hollow thin-walled truncated cone. The magnetic field generated by such VCS was considered by us in [8]. The height of the cones was chosen in the range of 500-1000 km in accordance with the results of [9]. Difference between the radius of the upper and lower bases was 100 km. The other parameters: a volumetric current density, a base radius, the center location coordinates, and the spatial orientation angles were determined separately by solving the inverse problem for each VCS. First, the inverse problem was solved for a not large region in the vicinity of the maximum of the *Z* component and then obtained parameters were refined during the iterative process. The method is described in detail in [9]. According to the results of [9], to separate correctly the systematic (dipole) and the nondipole part of MGF the seventh VCS has to be included in the model.The height of the corresponding truncated cone was 4200 km. The remaining parameters were obtained in the same way as for all other. The location and orientation of this VCS is not discussed in this paper, since the analysis of the systematic component is beyond the scope of the task.

**Results.**

The position and orientation in space of the cone-shaped VCSs were determined in the course of solving the inverse problem. The resultant magnetic field of the seven VCSs governs the spatial structure of the MGF at the distance of 10000 km from the Earth center. It should be noted that, despite a significant distance from the possible sources region, sevenVCSs turned out to be insufficient for a complete description of the MGF at that distance. The remainder of the approximation is also not discussed here. In the framework of this work, we limited ourselves to considering only world anomalies. An analysis of the location and orientation of the obtained VCSs relative to the heterogeneity of the structure of the lowest mantle (core-mantle boundary) are showed in Figs. 3-8. For all figures, the point of view is assumed to be located inside the sphere corresponding to the core-mantle boundary. The spatial distribution of the seismic wave velocity heterogeneities $\delta V$ is calculated according to the SAW642AN model at the core-mantle boundary and represented by the color on the spherical segment. The rotational displacement of this segment was chosen based on the best visual representation of the relative position of the VCS and the heterogeneities of the core-mantle boundary. XYZ axes correspond to the central coordinate system. For a better understanding how the spherical segment is oriented in space, arrows are added to figures which show the direction ofthe westwarddrift ofthe liquid outer core. One can readily see that the location of all VCSs is characterized by the presence of a high-speed mantle anomaly west of the VCS. These anomalies can form the core-mantle topography and obstruct the free western drift at the edge of the liquid core. And this involves an inception and buildup in the nondipole part. The VCS shown in Figs. 3 and 4 are practically "surrounded" by mantle anomalies of different intensities. These anomalies forming the topography hills can restrict incoming substance of liquid core to VCS, wherefore its magnetic moment has to be decreasing as it is observed.

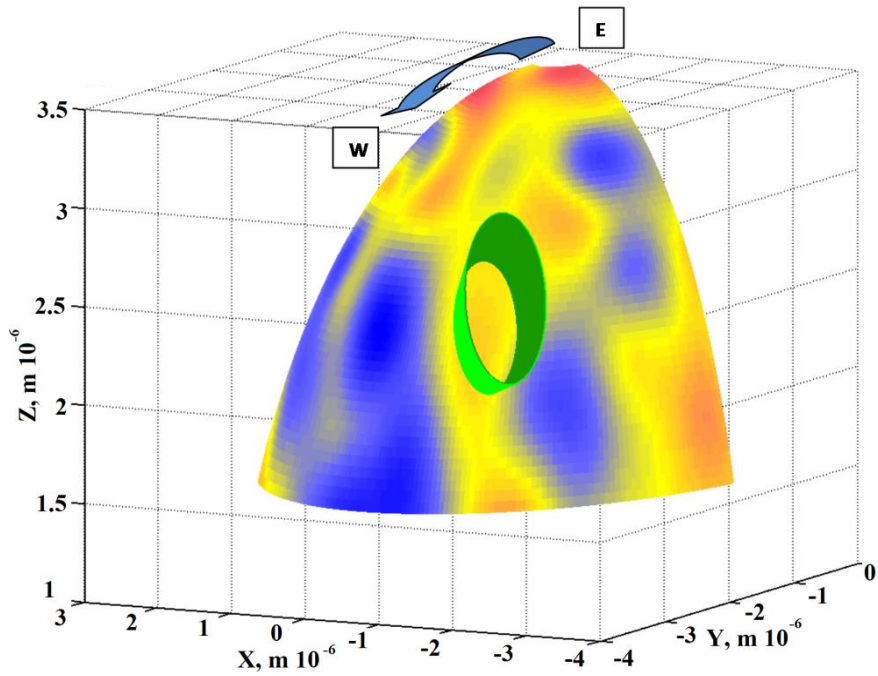

**Fig. 3** Location and orientation of the cone obtained for the first anomaly (Fig.1) relative the mantle structure at the core-mantle boundary. VTS is shown by green, $\delta V$ m/s is shown by color map.

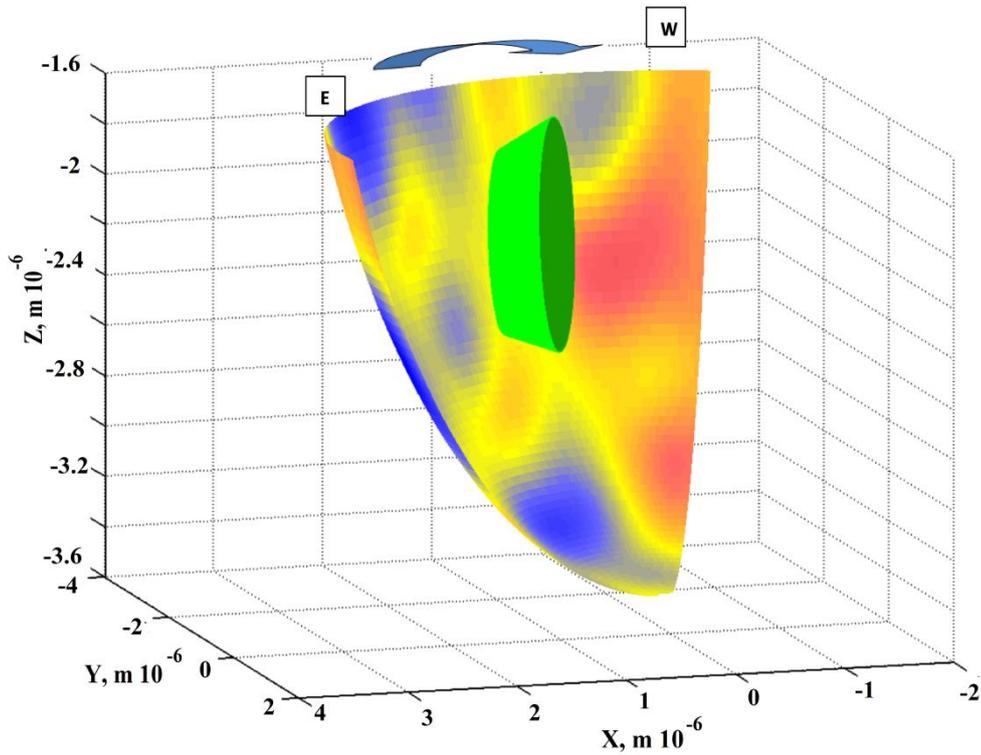

**Fig. 4** Location and orientation of the cone obtained for the fifth anomaly 5 (Fig. 1) relative the mantle structure at the core-mantle boundary. The legend is the same as in Fig. 3.

As for the 2nd anomaly, which is shown in Fig. 5, one can see that the mantle heterogeneities are mainly located west of VCSs. ThisVCS is characterized by a growing but the smallest magnetic momentum (MM). The corresponding MGF anomaly was actually described only at the very end of the 20th century [9]. Its MM growing, is connected with the position relative to the mantle heterogeneities. As it may be shownin Fig.5 these heterogeneities are placed west of VCS that allows an additional inflow of liquid core material.

The 3rd and the 6th VCS were obtained having the most powerful MM. Their location is shown in Figs. 6 and 7 correspondently. The mantle heterogeneities are also located west of VCSs. Thus, the westward drifting core material can result in growth of the VCS.

.

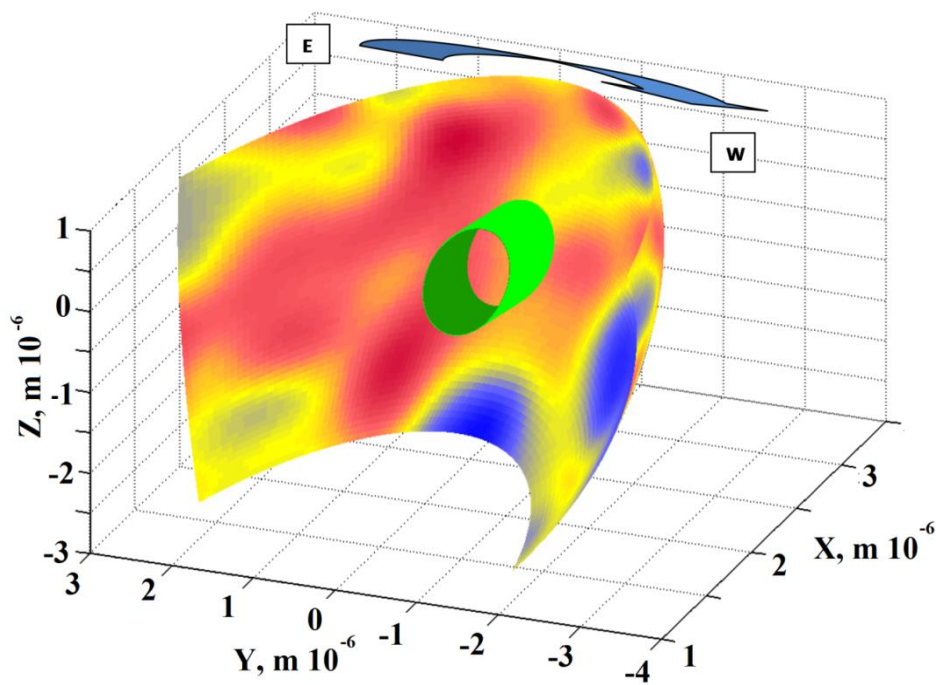

**Fig. 5** Location and orientation of the cone obtained for the second anomaly (Fig. 1) relative the mantle structure at the core-mantle boundary. Legend is the same as in Fig. 3.

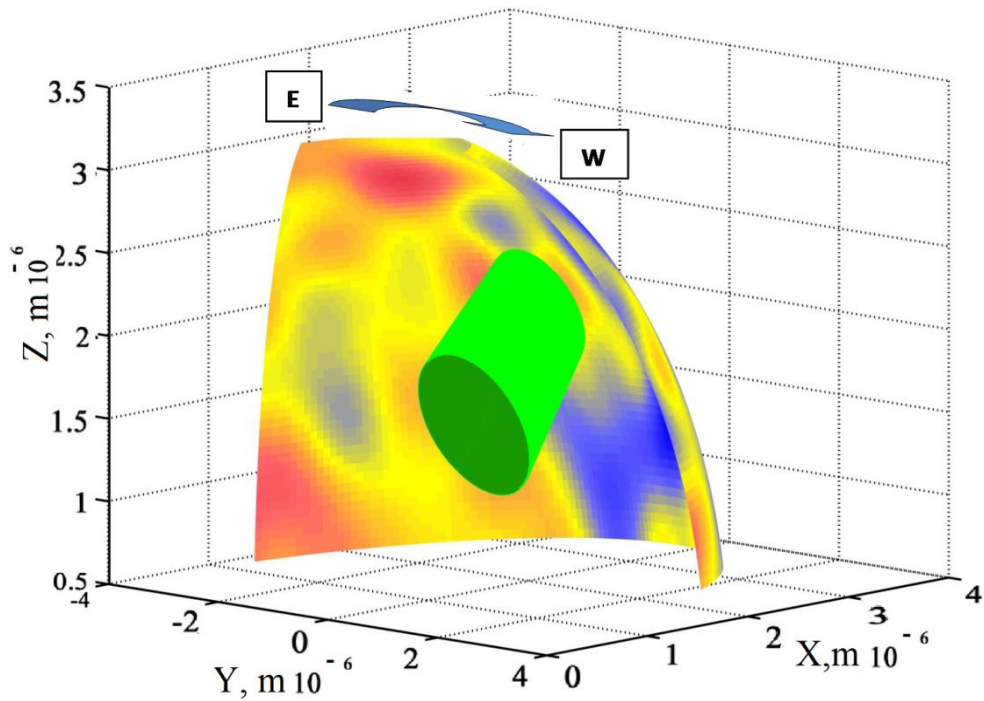

**Fig. 6** Location and orientation of the cone obtained for anomaly 3 (Fig.1) relative the mantle structure at the core-mantle boundary. The legend is the same as in Fig. 3.

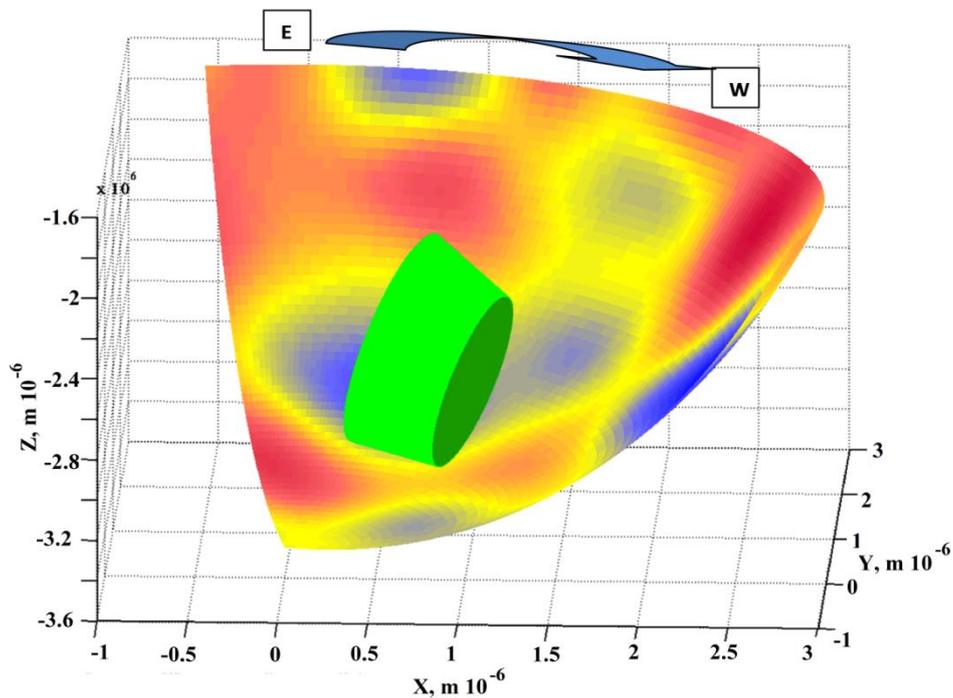

**Fig. 7** Location and orientation of the cone obtained for the sixth anomaly (Fig.1) relative the mantle structure at the core-mantle boundary. Legend is the same as in Fig. 3.

The VCS corresponding to the fourth anomaly (Fig.8) deserves a separate consideration. In [10], where to carry out the approximation by point sources, was obtained that the magnetic moment vector of source obtained for this anomaly tuned south-westward for a continuous period of 100 years. This is conceivably due to its displacing by the topography mantle heterogeneities.

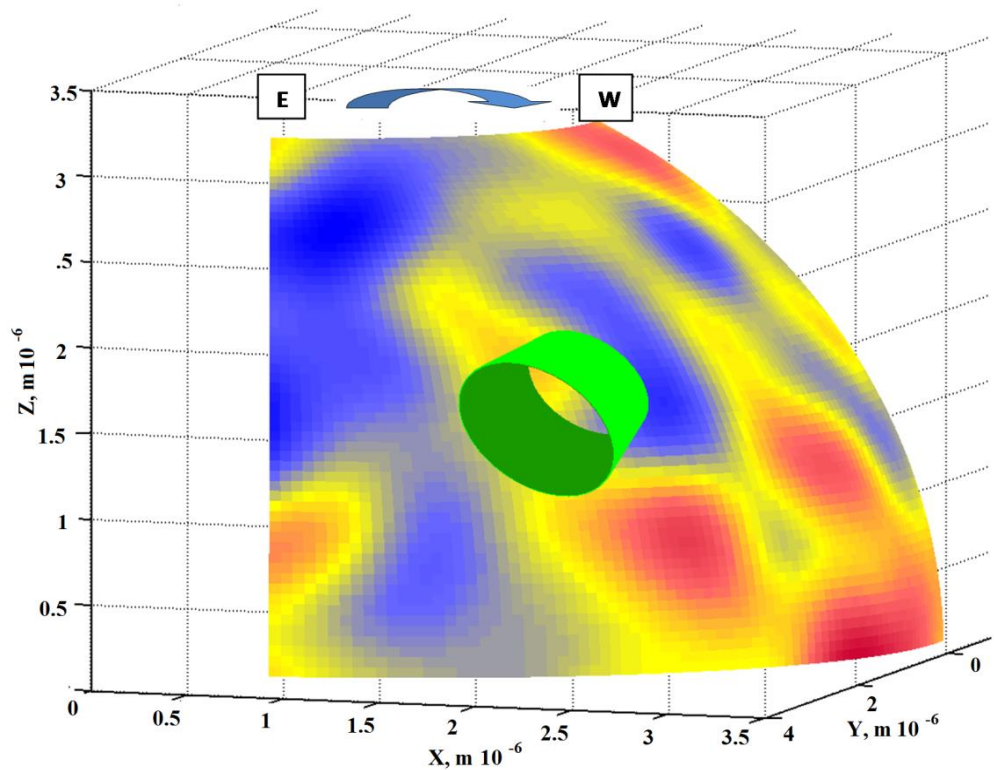

**Fig. 8** Location and orientation of the cone obtained for the fourth anomaly (Fig.1) relative the mantle structure at the core-mantle boundary. The legend is the same as in Fig. 3.

**Discussion.**

The main result obtained in this paper concerns the association between the structural heterogeneities of the lowest mantle and the origination of the world anomalies of the geomagnetic field and their secular variation. In fact, both problems are the subject of discussion. The presence of high-speed anomalies in the lowest mantle is confirmed by all modern global mantle models [11] and can be traced from subduction zones on the Earth's surface to the core-mantle boundary (CMB). The subducted part of lithospheric plates (slab) is colder and hence heavier relatively to the surrounding mantle. That result in so called slab pull. Finally the slab remains can penetrate into the lowest mantle. At larger scales, seismic tomography has become the primary tool for imaging the subducting slab in the lowest mantle. The geometry and the behavior of slabs varies not only among different subduction zones but also within the subduction zone[12].The ancient dense lithospheric slabs are concentrated into large agglomerations at the CMB.TheCMBtopography is poorly determined through seismology. Garcia &Souriau [14] found CMB topography of less than 4 km for waves with wavelengths longer than 300 km. Sze & van der Hilst [15] found the topography amplitudes of up to 5 km and reaching up to 13 km. But there is an overall trend towards assumed maximum amplitudes of around 1.5 km at long wavelengths [13]. So the authors [15] decided that this topography is unusually high and reduced amplitudes to 3 km. Authors of [16] have derived a model of CMB topography from the mantle dynamics. According to the authors this model should be useful at least at long wave lengths of several thousand kilometers with 8 km amplitude. From the other side the authors expect that the thermalboundary layer at the base of the mantle is on average around 300 km thick. In [17] it was obtained that the seismic wave anomalies in the uppermost 300 km of the outer core cannot be of thermal origin and should primarily reflect compositional heterogeneity.

Based on the above results, we assume that the denser slabs can partially penetrate below the CMB, generating the topographical "hills" on the CMB. This geometry impedes the free differential rotation of the liquid core substance, separating some part from the main generation cylinder and forming additional current structures. The comparative study carried out in this paper does not allow us to estimate the geometric dimensions of "hills". In this case, we are only talking about their effect on secular variations.

The thermal core-mantle coupling also affects the dynamics of the outer core. Cold slabs at the baseof the mantle are expected to increase the local heatflow. Several studies have been sought to interpret core flows obtained from secular variation of the MGF in terms of regional variations in heat flow [18, 19, 20]. Numerical simulations reveal a tendency to lock the pattern of convection to the pattern of time-dependent heat flow at the CMB. Similar conclusions are drawn from numerical geodynamo models [10, 21, 22]. Non-homogeneous boundary conditions in geodynamo models yield persistent structure in the time-averaged flow [21, 22, 23], although there can be substantial variation about the average. A solution to the full dynamo equations with lateral variations in heat flux on the outer boundary defined by the shear wave velocity of the lowermost mantle is presented in [24]. The assumption was that cold regions in the lower mantle could cause preferential cooling of the core, downwelling, and concentration of radialmagnetic flux at the core surface. As a result the four main equatorially symmetric flux lobes were obtained in magnetic field at the CMB. Authors strongly suggest that geomagnetic field morphology is dominated not only by geometry related to the inner core but also by the seismically fast structure in the bottom few hundred kilometers. Let us note that the authors have taken into account only thermal variations as having a dominant influence and have not considered compositional variations. In addition, a very averaged model of the velocity heterogeneity near the CMB was used in simulations.Therefore we are not looking for absolute matches of our results with those obtained in geodynamo models. Nevertheless , they do not contradict each other.

**Conclusion**
The results obtained in this paper provide implicitly support for our hypothesis that the topography of the CMB has a more complex relief than is suspected. The cold regions of the lowermost mantle associated with the higher density slabs can penetrate the liquid core, forming topographic "mountains". These heterogeneities prevent the free differential rotation of the liquid core relative to the mantle, what leads to a partial outflow of liquid core material from the main generation cylinder, the formation and development of new eddies of different scales. These two processes lead, on the one hand, to a decrease in the systematic component of the MGF and on the other hand, to an increase in its nondipole part.
The formation of such "mountains" at the core-mantle boundary is possible due to processes that took place many millions of years ago, when significant volumes of the earth's crust were absorbed in ancient zones of subduction [10]. Due to the different speeds of tectonic processes in the past, the topography of the core-mantle boundary changed in time. A significant increase in the size of such heterogeneities could lead to the destruction of the main generation volume — excursions and inversions.